\documentclass[sigconf]{acmart}

\usepackage{booktabs} 
\acmConference[Preprint version]{Preprint version}
\acmYear{2023}
\copyrightyear{2023}

\begin{document}
\title{Towards an Ontology-Driven Approach\\ for Process-Aware Risk Propagation}
  

\author{Gal Engelberg}
\affiliation{%
  \institution{Accenture Labs, University of Haifa}
  \streetaddress{P.O. Box 1212}
  \city{Haifa} 
  \country{Israel} 
}
\email{gal.engelberg@accenture.com}

\author{Mattia Fumagalli}
\affiliation{%
  \institution{Free University of Bozen-Bolzano}
  \city{Bolzano} 
  \country{Italy} 
}
\email{mattia.fumagalli@unibz.it}

\author{Adrian Kuboszek}
\affiliation{%
  \institution{Avanade}
  \country{Poland}}
\email{a.kuboszek@avanade.com}


\author{Dan Klein}
\affiliation{%
  \institution{Accenture Labs}
  \country{Israel}}
  \email{dan.klein@accenture.com}


\author{Pnina Soffer} 
\affiliation{%
 \institution{University of Haifa}
 \city{Haifa} 
 \country{Israel}}
\email{spnina@is.haifa.ac.il}
 
\author{Giancarlo Guizzardi}
\affiliation{%
  \institution{University of Twente}
  \country{The Netherlands}
}
\email{g.guizzardi@utwente.nl}

\begin{abstract}
The rapid development of cyber-physical systems creates an increasing demand for a general approach to risk, especially considering how physical and digital components affect the processes of the system itself. In risk analytics and management, \textit{risk propagation} is a central technique, which allows the calculation of the cascading effect of risk within a system and supports risk mitigation activities. However, one open challenge is to devise a \textit{process-aware} risk propagation solution that can be used to assess the impact of risk at different levels of abstraction, accounting for actors, processes, physical-digital objects, and their interrelations. To address this challenge, we propose a process-aware risk propagation approach that builds on two main components: \textit{i.} an \textit{ontology}, which supports functionalities typical of \textit{Semantic Web technologies (SWT)} and semantics-based intelligent systems, representing a system with processes and objects having different levels of abstraction, and \textit{ii.} a method to calculate the propagation of risk within the given system. We implemented our approach in a proof-of-concept tool, which was validated and demonstrated in the cybersecurity domain.
\end{abstract}

%
%

\ccsdesc[500]{Information systems~Decision support systems}
\ccsdesc[300]{Risk ~Assessment and propagation}

\keywords{Risk propagation, risk assessment, ontology-driven risk propagation, risk, ontology}

\maketitle


\section{Introduction}

Risk is a pervasive phenomenon, depending on events that occur in a connected world, where objects interact with each other and cannot be taken in isolation. This structural aspect of risk-affected environments motivates the large and successful application of graph algorithms for analyzing how risk spreads in a given system. The application of graph algorithms to assess the risk spreading level in a system is commonly known as risk propagation \cite{jiang2016identifying}. Typically, risk propagation approaches are used in risk analytics and management to calculate the cascading effect of risk within a network of nodes representing a system, and are aimed at supporting risk identification, quantification, and mitigation activities.   

At the current state, risk propagation techniques are applied in different domains where processes play a central role. For instance, risk propagation is broadly adopted to analyze how occurrences of risk affect the sustainability of producer-consumer networks in supply chains \cite{choudhary2022risk}. Similarly, the propagation of risk is used to assess the impact of cyber-attacks on different assets of a given system \cite{kavallieratos2021cyber}.  
In this context, it has been widely recognized that one key open challenge is to devise a risk propagation solution that can be used to measure the cascading effect of risk in systems that involve dependencies between processes and physical objects \cite{gonzalez2021quantifying,engelberg2022process}. For instance, \textit{how can cybersecurity risk be propagated from a cyber infrastructure to the business processes of an organization? How may a machine breakdown affect the productivity of a company?} \textit{How may lead-time variability risk  affect a supply chain or a manufacturing environment?} \textit{How can we quantify the risk of machinery energy consumption deviation from the allowed thresholds and propagate the risk to the business processes of the host organization?}  

\begin{figure*}[t]
  \centering
  \includegraphics[width=1\linewidth]{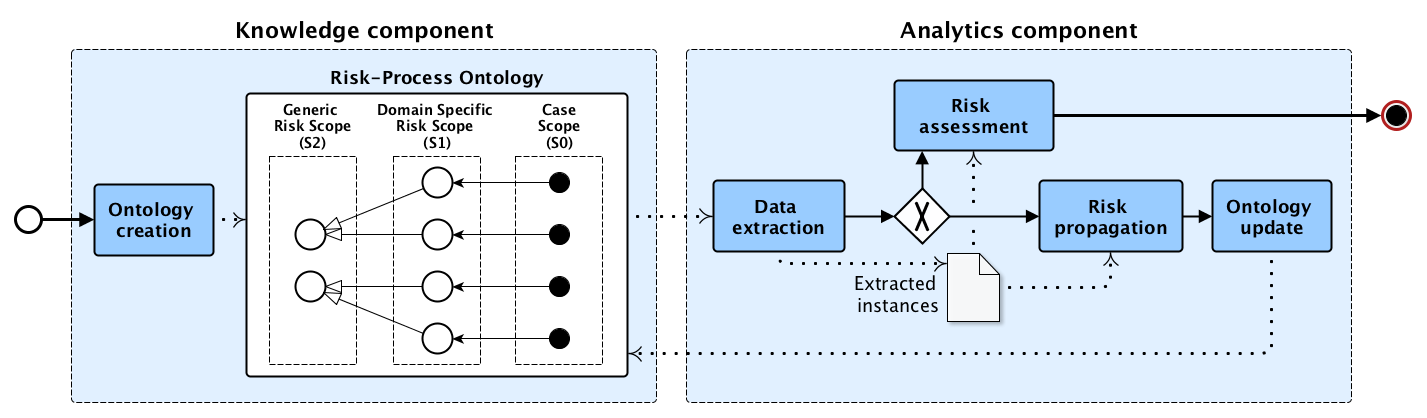}
  \caption{Process-aware risk propagation approach: overall view.}
  \label{approach}
\end{figure*}

All the above challenges can benefit from a process-aware\footnote{Here, by adapting the definition provided in \cite{dumas2013fundamentals}, we take ``process-aware'' as \textit{``regarding systems that involve processes''}.} approach to achieve better risk propagation. Such an approach should be able to leverage knowledge about how different processes, objects, and activities connect with each other, in domain-specific contexts (e.g., \textit{customer relationships}, \textit{enterprise planning}, \textit{cyber assets}, and \textit{supply chain}), and also at a domain-agnostic level, by covering concepts that are always present in different application contexts. 

This paper advances the state-of-the-art in the research of risk-propagation techniques, by proposing a process-aware approach that is aimed at facilitating the assessment of risk propagation between processes and objects with different levels of abstraction. The contribution leverages the combination of \textit{i.} an ontology, which supports functionalities typical of \textit{Semantic Web technologies (SWT)}\footnote{\url{https://www.w3.org/standards/semanticweb/}} and semantics-based intelligent systems, encoding a set of rules to be used for representing the risk dependencies within a system composed of objects and processes, and \textit{ii.} a method to calculate the propagation of risk within the represented system. We implemented our approach in a proof-of-concept tool, which was validated in the cyber-security domain. 

The remainder of this paper is structured as follows. Section \ref{req} lists the requirements that drove the design of our approach. Section \ref{approachS} describes the method embedded in our approach. In Section \ref{demoS} we discuss some implementation aspects and we report on a demonstration to validate our solution. Here we also discuss some implications and limitations of the current implementation. Then, in Section \ref{rel} we situate our contribution with respect to related work. Finally, in Section \ref{cons} we reflect on our results and elaborate on future work.

\section{Requirements}
\label{req}
Following the \textit{design science paradigm} \cite{hevner2004design}, we grounded the design of the proposed approach on a preliminary \textit{problem identification} activity. In this phase, we gathered feedback from 5 target users, namely cyber-security experts, who have been involved in risk assessment activities and that have been working on the identification of risk causes and risk dependencies between business processes and physical objects. We ran open-ended interviews and the main open questions we asked were about: \textit{i.} \textit{the relevance of an approach for facilitating the assessment of risk and its propagation at different levels of abstraction} and \textit{ii.} \textit{what is required to facilitate the assessment of risk and its propagation}. This preliminary step helped us in improving our awareness of the problem, better understanding the related work, and better identifying the features that our solution should offer to the end-users. Experts' feedback was also pivotal in the definition of a set of \textit{functional} (i.e., \textit{qualitative}) requirements, which are needed to design the approach and evaluate the artifact in which our contribution is embedded. 

\begin{figure*}[t]
  \centering
  \includegraphics[width=0.7\linewidth]{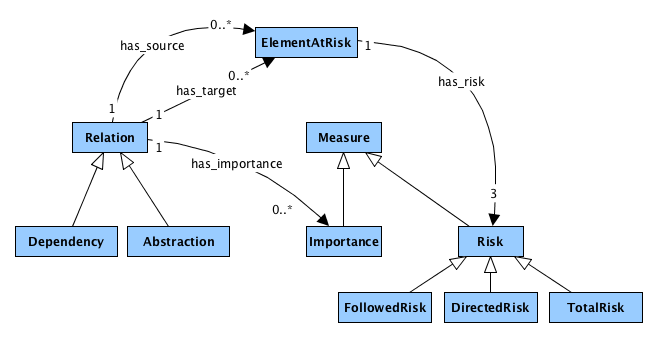}
  \caption{$S_2$ concepts and relations.}
  \label{model}
\end{figure*}

We mapped the key features that specify \textit{what} our approach should do into the following \textit{functional requirements}:

\begin{itemize}
    \item[\textbf{R1.}] The approach should be able to facilitate the task of propagating a risk that was measured at the \textit{physical level of a system} (e.g., a machine breakdown), towards its \textit{business process abstract level} (e.g., company productivity). This involves the capability to investigate how an attacker can compromise both the infrastructure assets and the business process goals of an organization.  
    \item[\textbf{R2.}] The approach should allow the users to easily assess risk and, in particular, to easily access the risk propagation output, possibly having visualization support to browse and analyze the data. This means also being able to filter out parts of the output according to \textit{ad hoc} queries. Notice that from the interviews came out that this requirement involves also the risk propagation and quantification method being explainable and understandable by domain experts.  
    \item[\textbf{R3.}] The approach should be able to support the users in identifying the root causes of risk, prioritizing the mitigation activities, and suggesting a relevant remediation plan. This will involve, also, for instance, the possibility to consider different business objectives for the risk mitigation task (e.g., reducing the risk for a single activity, a single process, a production line, or a factory).
    \item[\textbf{R4.}] The approach should be able to keep track of risk propagation over time. For instance, how does the propagation change after applying a mitigation step? For the same processes and objects, are there any different risk propagation phenomena at different times?  
    \item[\textbf{R5.}] The approach should allow discovering if some elements are at risk even if they are not ``directly'' connected. This should happen by considering different types of relations between elements at risk, like causal dependencies or physical connections. The goal of this requirement is to enable the propagation of risk when an element is a part, for instance, of a causal chain, or simply a component of a device.
\end{itemize}

\section{Approach Description}
\label{approachS}

The approach we propose is grounded on the standard definition of risk provided in \cite{ISO2018}. Accordingly, we use risk to ``\textit{quantify the possibility of reaching some given objectives}'', where such a quantity value is derived from the combination of the probability that a certain \textit{risk event} occurs (as a \textit{perturbation} of the plan for reaching the objectives) and a set of ``\textit{severity values}''. For example, suppose that an attacker has read/write access to a database, namely, he can damage the database \textit{integrity} and \textit{confidentiality}. The read/write access represents the risk event and the severity values will be associated with the database integrity and confidentiality features.

We employ here a simplified definition of risk as ``an effect of uncertainty on objectives''. Risk is a polysemic term, which covers multiple phenomena including risk magnitude, risk assessment, vulnerability, loss and threat events, etc. Moreover, it can be better represented as a \textit{relational property} (e.g., something is ``at risk'' w.r.t. to a particular goal and in a particular context, e.g., one can be at the same time at risk of missing a flight, at risk of contracting COVID-19, etc.). An in-depth analysis and formalization of the risk notion with regards to the proposed approach (by also leveraging previous work as in \cite{sales2018common,oliveiraontology}), is part of the immediate future work.

In the scope of this paper, the main observation is that we calculate the risk as $R = P\ast(S_1, ..., S_n)$, where $P$ provides the probability that a risk event occurs, and each $S_j$ encodes a severity value. In the context of our solution, the propagation task will start from a given risk value, associated to a given ``\textit{risk event}'' (e.g., ``\textit{device damaging}''). The whole approach is aimed at capturing how the risk associated with this risk event can spread through all the elements (objects and processes) involved (directly or indirectly) in the event itself.

Figure \ref{approach} provides an overall view of our solution, composed of a \textit{knowledge component} and an \textit{analytics component}.  

\begin{figure*}[t]
  \centering
  \includegraphics[width=1\linewidth]{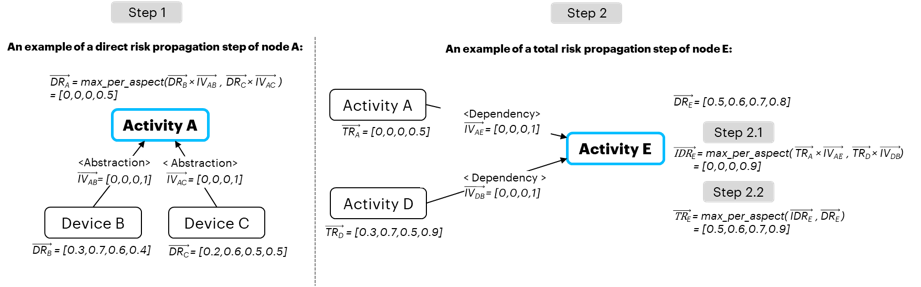}
  \caption{Risk propagation steps: running example.}
  \label{propex}
\end{figure*}

\subsection{Knowledge Component}

The knowledge component holds an ontology (denoted in Figure \ref{approach} as \textit{Risk-Process Ontology}), which is in turn divided into three scopes. The \textit{first scope}, what we call here $S_2$, is composed of a set of generic concepts and relations related to Risk, which are always required independently of any specific business domain. The \textit{second scope}, $S_1$, extends $S_2$ with a set of domain-specific concepts and relations. $S_1$, is then mapped into the \textit{third scope} $S_0$, composed of a use-case-specific types and instances.

Figure \ref{model} provides a lightweight representation of $S_2$, composed of the minimal set of constructs required for the process-aware risk propagation task. The main concept in this scope is \texttt{ElementAtRisk} which stands for both process types, or objects at risk. For example, an \texttt{ElementAtRisk} could be specialized in $S_1$ by a concept representing a physical component of a system, such as a \textit{``machine''}, or a business abstract concept such as a \textit{``business activity''}. We keep implicit the different types of \texttt{ElementAtRisk} (as objects, process types) including their interrelations, and that the ultimate scope is with regards to business objectives, and values. However, since here the main goal is to propose the overall approach, we take this lightweight model, which will be extended in future work.  

Within a system of \texttt{ElementAtRisk}, the risk is propagated from one element to another according to their relations. Within the current approach, in order to model risk propagation, we identified two main types of relations. First, \texttt{Dependency} relations, which are mainly used to model phenomena where the risk is propagated through a workflow composed of processes. For instance, two business activities can be connected by \texttt{Dependency} relations like \textit{``triggers''} or \textit{``causes''}. Second,  \texttt{Abstraction} relations, represent cases where the risk is propagated from a lower to a higher level of abstraction. For example, the risk of a physical machine can be propagated to related business activities. Given a network of elements at risk and their connections, we identify three types of \texttt{Risk}. We call \texttt{FollowedRisk} the risk propagated through \texttt{Dependency} relations and \texttt{DirectedRisk} the risk propagated through \texttt{Abstraction} relations. \texttt{TotalRisk}, in turn, stands for the overall risk of an object, considering both its \texttt{DirectedRisk} and \texttt{FollowedRisk}. 

Notice that the knowledge component is aimed at supporting risk calculation from different perspectives, which can be represented within $S_2$ through the \texttt{Measure} concept and some \textit{ad hoc} attributes. For example, in a cybersecurity use case, the risk could be quantified as from the CIA-triad standard \cite{fenrich2008securing}, namely according to its potential impact on \textit{availability}, \textit{confidentiality}, and \textit{integrity} of the related business activities; while in sustainability use-case, one can propagate the risk of a carbon-footprint just focusing on the \textit{deviation} from the machine level to the process level. Finally, the knowledge component allows the user to control the \textit{amount of risk propagated from one element to another} via the \texttt{Importance} concept, which is used to weight any given relation. For example, a confidentiality risk that was measured over a device and propagated to its correlated business activity should not necessarily be propagated to the following activity. In that case, the system supports omitting the propagation of a confidentiality risk from an activity to the following by setting an \texttt{Importance} of zero.
\subsection{Analytics Component}

\begin{figure*}[t]
  \centering
  \includegraphics[width=0.6\linewidth]{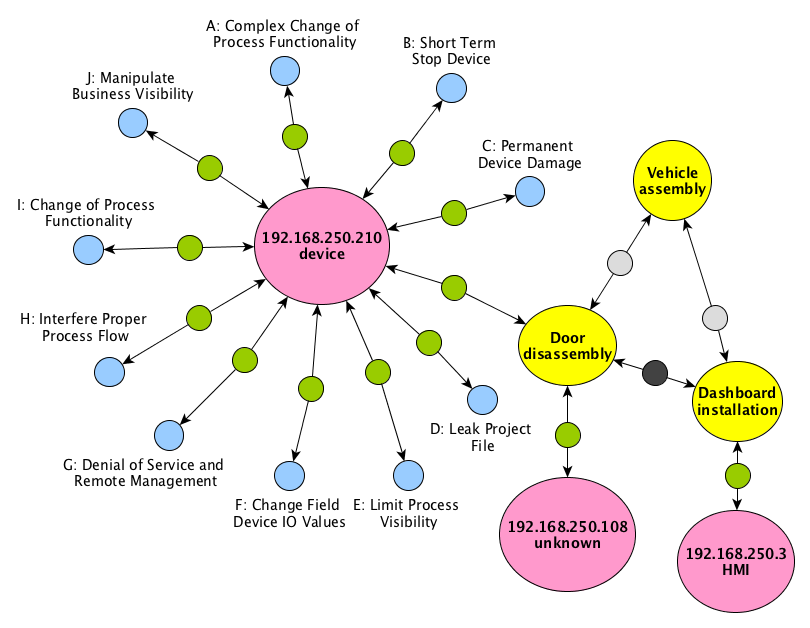}
  \caption{A subgraph of the case scope ($S_0$).}
  \label{subgraph}
\end{figure*}

Once the Risk-Process ontology is set, the analytics component is used for a data extraction step that consists of querying the ontology through the $S_2$ constructs. The data extraction step returns a \textit{labeled property graph structure} \cite{das2014tale} where each node represents an instance of an \texttt{Element\-AtRisk} and each edge represents an instance of a \texttt{Relation}. The \texttt{Risk} and \texttt{Importance} values are then represented as vectorized properties of nodes and relationships respectively. Notice that, the proposed approach assumes that the risk over the leaf nodes (elements with a lower level of abstraction) is given prior to the risk propagation task. 

Once the labeled property graph is generated, risk propagation can be performed. This task is performed in two steps, where, at each step, the graph is traversed via a \textit{Depth-first Search (DFS)} algorithm \cite{awerbuch1985new}. A risk propagation for a single node is in turn defined according to a risk function denoted as \textit{max\_per\_aspect}. In the proposed method we take a worst-case scenario approach by quantifying the risk according to the \textit{maximal risk per perspective}. For example, in a case where a business activity depends on two devices, and each has a different availability risk. A worst-case scenario approach assumes that both devices could be compromised by an attacker, and a shutdown of at least one device will disable the correlated activity. Thus, the propagated risk towards the business activity is set according to the maximal availability risk of both devices. The risk function gets a bag of vectors ordered by the different risk perspectives and returns the maximal value for each perspective. Notice that, in the future, we plan to support multiple risk functions set in $S_2$.

Figure \ref{propex} provides an example of a risk propagation task and the two steps of which it is composed. The risk in the running example is a vector composed of four values, where each value represents a risk quantification from a different perspective. For example, in a cyberattack the following device perspectives could be affected: \textit{confidentiality}, \textit{integrity}, \textit{safety}, and \textit{availability}. 

\begin{itemize}
    \item [1] In the first step of the risk propagation task, the \texttt{DirectedRisk} (see $\overrightarrow{DR}$ in Fig. \ref{propex}, Step 1) is propagated from the leaf nodes to the nodes with a higher level of abstraction using the \texttt{Abstraction} relation. The bag of vectors for each node is composed of the \texttt{DirectedRisk} vectors of the incoming nodes ($\overrightarrow{DR}_B$ and $\overrightarrow{DR}_C$), multiplied by the corresponding \texttt{Importance} vectors ($\overrightarrow{IV}_{AB}$ and $\overrightarrow{IV}_{AC}$) over the incoming edges. The multiplication is an \textit{element wise}, namely, each element in the \texttt{DirectedRisk} vector is multiplied with the corresponding element in the \texttt{Importance} vector.  
    \item [2] Once the \texttt{DirectedRisk} is propagated across the graph, the second step occurs according to two main sub-steps:
    \begin{itemize}
        \item [2.1] The \texttt{FollowedRisk} vector of a node (denoted as $\overrightarrow{IDR}$) is calculated. In this case, the bag of vectors for each node is composed of the \texttt{Total\-Risk} vectors (denoted as $\overrightarrow{TR}$) of its incoming nodes multiplied by the corresponding \texttt{Importance} vectors over the incoming edges. Notice that, the \texttt{Followed\-Risk} vector over leaf nodes is set to zero.
        \item [2.2] The \texttt{Total\-Risk} of a node is calculated. In that case, the bag of vectors is composed of its \texttt{Followed\-Risk} and \texttt{Directed\-Risk} vectors. Once the risk propagation task is concluded, the results are updated in the Risk-Process ontology.
    \end{itemize}
\end{itemize}

Finally, the analytics component also accounts for another step, what we call here ``risk assessment''. Here, the ontology can be queried to assess and analyze the risk state of the whole system, namely the risk of the \texttt{ElementAtRisk} with the highest level of abstraction. Furthermore, through the risk assessment step, it is possible to return an alert considering the deviation of the quantified risk from a pre-defined threshold (denoted as a \textit{cardinal risk}). Similarly, an analyst could use this step to analyze what is the element at cardinal risk, identify the risk's root causes, and prioritize mitigation steps accordingly. Still, the proposed approach enables the detection of elements at risk, even if their directly connected elements are not at risk. For example, a manual activity could be affected by a cyberattack since it is followed by another activity that depends on a device at risk of being compromised. Similarly, this type of propagation is applicable in recently Log4j\footnote{\url{https://logging.apache.org/log4j/2.x/}} supply-chain attack.

\section{Implementation and Demonstration}
\label{demoS}

This section discusses implementation details, and reports on a demonstration to validate the proposed approach. 

\vspace{0.5em}
\noindent\textbf{Implementation}. The \textit{knowledge component} is deployed on \textit{Neo4J graph database platform}\footnote{\url{https://neo4j.com/}}, the \textit{analytics component} and the whole pipeline orchestration are implemented as a \textit{Python}\footnote{\url{https://www.python.org/}} application which interacts with Neo4J via an \textit{ad hoc} Neo4J python library\footnote{\url{https://neo4j.com/docs/api/python-driver/current/}}. The program and the database interact at three main stages, as described in Figure \ref{approach}. First, the import of the ontology into the database (denoted as \textit{Ontology creation}). Second, the export from the database into the program memory towards the risk propagation task (denoted as \textit{Data extraction}). Thirdly, the database update with the risk propagation results (denoted as \textit{Ontology update}). 

To represent the ontology scopes, we adopted the \textit{Ontology Web Language (OWL)}\footnote{\url{https://www.w3.org/OWL/}}. The model's \textit{concepts}, \textit{relations}, and \textit{attributes} are expressed as \textit{classes}, \textit{object properties}, and \textit{data properties}, respectively. $S_0$ is expressed as classes' \textit{individuals} and their \textit{properties} assertions. Once the OWL file of the three scopes is constructed, we import it to the database using the Neo4J \textit{NeoSemantics} plugin\footnote{\url{https://neo4j.com/labs/neosemantics/}}. This plugin takes a \textit{Resource Description Framework (RDF)}\footnote{\url{https://www.w3.org/RDF/}} structure and transforms it into a \textit{Labeled Property Graph (LPG)} structure. In this structure, the constructs of the model and the data are represented as nodes and edges within a graph database.

\begin{table*}[]
\scriptsize
\centering
\includegraphics[width=0.7\linewidth]{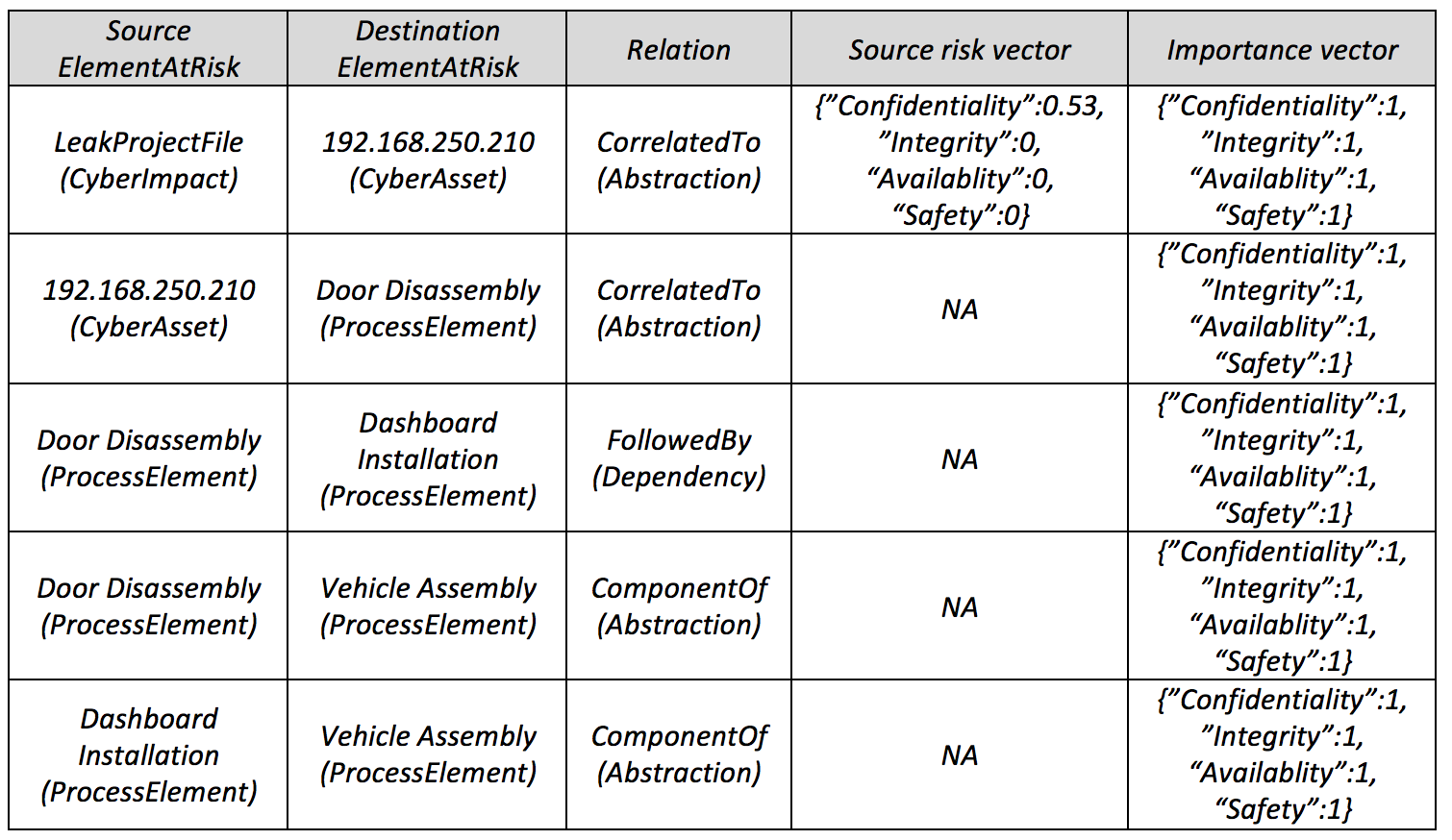}
\caption{An example of data extraction output.}
\label{tab0}
\vspace{-3em}
\end{table*}

\begin{table}[]
\centering
\includegraphics[width=0.75\linewidth]{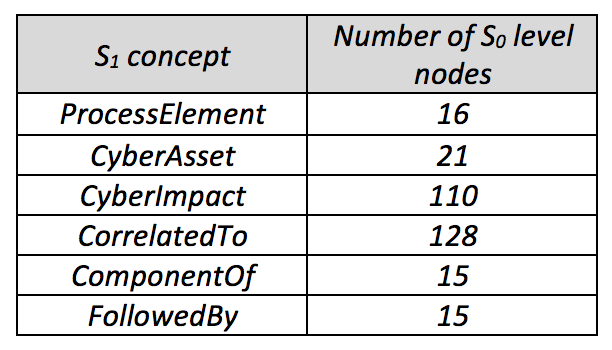}
\vspace{0.5em}
 \caption{\textit{Number of case scope $S_{0}$ nodes per $S_{1}$ risk scope concept}}
  \label{tab1}
  \vspace{-3em}
\end{table}

\vspace{0.5em}
\noindent\textbf{Demonstration}. We demonstrate the approach through a cybersecurity risk assessment use case of a vehicle assembly manufacturing process. This example serves for showing how the proposed approach can be used for quantifying the risk of devices being compromised by a cyberattack, and then measuring the impact over the domain-specific risk scope. 

In this scenario, the main concepts captured by the ontology (see $S_2$ and $S_1$) can be grouped into \textit{a)} a \textit{physical layer} composed of devices (denoted as \texttt{Cyber\-Asset}) that could be compromised by an attacker; \textit{b)} potential \textit{intervention actions} (denoted as \texttt{Cyber\-Impact}), which an attacker could perform over each device; \textit{c)} processes (each one grouped as as \texttt{Process\-Element}). Cyber assets and process/activity elements are connected via relations of type \texttt{Correlated\-To}, process/activity elements are connected via relations of type \texttt{Component\-Of} and \texttt{FollowedBy}. According to the $S_2$ distinctions, \texttt{Correlated\-To} and \texttt{Component\-Of} are classified as \texttt{Abstraction} relations, \texttt{Followed\-By} is classified as a \texttt{Dependency} relation. Considering the given conceptualization, the risk is then measured over the different \texttt{Cyber\-Impact} instances and propagated to \texttt{Cyber\-Asset} and \texttt{Process\-Element} instances. Notice that, in this demonstration we measured risk according to the commonly used \textit{CIA-triad} for a cybersecurity risk assessment, where the risk vector is composed of the perspectives of \textit{confidentiality}, \textit{integrity}, and \textit{availability}. For example, a denial-of-service \texttt{Cyber\-Impact} holds a substantial risk of availability, while a data manipulation \texttt{Cyber\-Impact} holds a substantial risk of integrity and confidentiality. Since the risk is measured within an industrial facility, we extend the standard approach with a safety perspective.

Figure \ref{subgraph} shows a snapshot of the case scope ($S_0$) instantiating the ontology concepts. And Table \ref{tab0} shows an example of the risk propagation output of the represented graph. The \texttt{Followed\-By} (denoted as light-grey nodes) and the \texttt{Component\-Of} (denoted as dark-grey nodes) represent relations between process elements (denoted as yellow nodes), which are represented by three processes, namely: \texttt{Vehicle\-Assembly}, \texttt{Door\-Disas\-sembly} and \texttt{Dashboard\-Installation}. The graph provides then the common cyber assets for each \textit{Process\-Element} as well. This is represented by the \textit{Correlated\-To} relation (denoted as green nodes) between process elements and cyber assets instances (denoted as pink nodes). As from Figure \ref{subgraph} \texttt{Door\-Disassembly} is connected with two \texttt{Cyber\-Asset} instances, and \texttt{Dashboard\-Installation} relates to just one instance. Finally, the graph encodes the potential vulnerabilities of the selected cyber assets, by connecting them to a set of threat instances, categorized as \texttt{CyberImpact} (denoted as blue nodes), and each one associated with a given risk vector\footnote{Notice that each risk vector was derived by leveraging an \textit{ad hoc} analysis step similar to \cite{kerschbaum_assessing_2018, hadar_cyber_2020}. A detailed explanation of this step and the multiple options that can be adopted is out of the scope of this contribution.}. The subgraph in Figure \ref{subgraph} shows also that one CyberAsset instance (far left) is connected to 10 \texttt{Cyber\-Impact} instances, while the rest of the \texttt{Cyber\-Asset} instances in the subgraph are not connected, i.e., they can be considered as ``secure''. 

Once the ontology is set and imported to Neo4J, we run the \textit{data extraction} step. Here, we use a \textit{cypher}\footnote{\url{https://neo4j.com/developer/cypher/}} query to extract elements at risk and the relations that are relevant to the risk propagation task. The query uses $S_2$ constructs to support various domain-specific entities and relations and returns a set of records encoding a \textit{i)} relation between a source to a destination element, \textit{ii)} the risk vector over the source object, and the \textit{iii)} importance vector of the relation. 

\begin{table*}[h!]
  \centering
  \includegraphics[width=0.9\linewidth]{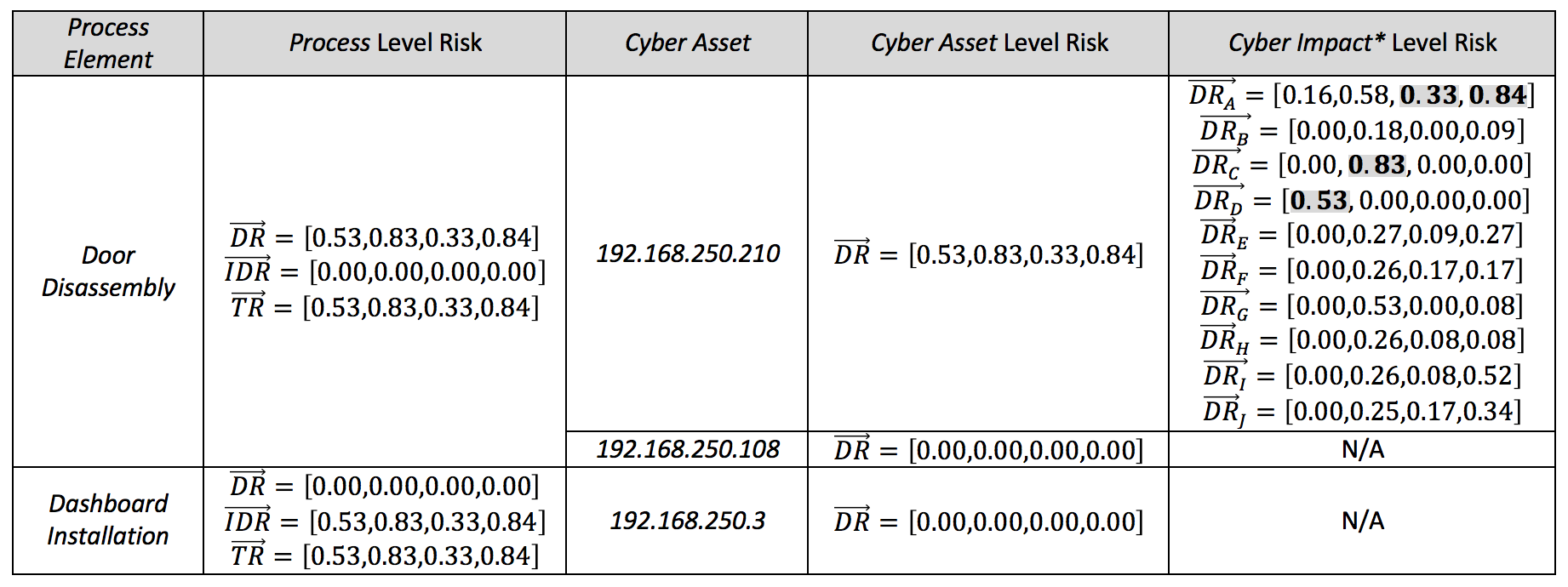}
  \caption{Risk propagation output example. A: ``Complex Change of Process Functionality'', B: ``Short Term Stop Device'', C: ``Permanent Device Damage'', D: ``Leak Project File'', E: ``Limit Process Visibility'', F: ``Change Field Device IO Values'', G: ``Denial of Service and Remote Management'', H: ``Interfere Proper Process Flow'', I: ``Change of Process Functionality'', J: ``Manipulate Business Visibility''.
}
  \label{output}
\end{table*}

Table \ref{tab1} reports the output of this step, providing the number of nodes for each domain-specific risk scope concept. Notice that, for clarity, we used here an importance vector of one for all the relation instances. In that stage, the risk is associated only with \texttt{CyberImpact} objects, namely, with the leaf nodes of the extracted graph.

Finally, the data shown in Tab. \ref{output} describes the output of the risk propagation task, from the \texttt{CyberImpact} objects to the \texttt{Cyber\-Asset} objects, then to the \texttt{Process\-Element} objects given the $S_0$ data extracted from the designed ontology. The far-right column encodes the risk vectors for the \texttt{CyberImpact} instances (i.e., the leaf blue nodes in the example of Fig. \ref{subgraph}), where we labeled each instance with a common potential threat in the cyber security context. For instance, a device may stop for a certain period of time (B), or a device may be damaged (C). Such a \texttt{CyberImpact} labeling was grounded on previous research work analysis on attack techniques for industrial control systems \cite{alexander2020mitre}. 


\vspace{0.5em}
\noindent\textbf{Discussion}. Given the above demonstration, we see four main observations. The first is that the combination of the generic, domain-specific, and case risk scope offers a solution to support \textbf{R1} (see Section \ref{req}). These three scopes together cover, indeed, physical and business process level concepts. Notice that, even if the demonstration focuses on the specific cyber-security case, this does not prevent the application of the approach over different domains (which can be encoded by the domain-specific risk scope). Secondly, the risk assessment step enables the user to browse, query, and analyze the input/output graphs by leveraging the Neo4J functionalities, thus facilitating the user in all the risk assessment activities, and fostering understandability, as required by \textbf{R2}. Thirdly, by adopting the maximal risk per perspective, the main causes of risk can be straightforwardly derived as a consequence of the propagation process, thus addressing \textbf{R3}. Looking, for instance, at the \texttt{Process\-Element} nodes in Table \ref{output}, we can see that the ``DoorDisassembly'' \texttt{Process\-Element} has a \texttt{DirectedRisk} which is caused by three \texttt{CyberImpact} instances over the ``192.168.250.210'' \texttt{Cyber\-Asset} (denoted as ``Complex Change of Process Functionality'' (A), ``Permanent Device Damage'' (C), ``Leak Project File'' (D)). Notice that in this case, we have three risk causes, because of the \textit{maximal risk per perspective} we adopted, where, given a list of vectors, we make one vector out of the maximum values (highlighted in grey in Table \ref{output}) for each index across all values. Differently, for what concerns \textbf{R4}, the proposed implementation still needs to be extended. In the current state, indeed, it is not possible to keep track and combine the multiple risk assessment output over time, thus limiting the analysis at different temporal snapshots. Finally, through the exploitation of dependencies and abstraction relations, the approach is able to uncover ``implicit'' (followed) or ``direct''/``explicit'' risk (see \textbf{R5}). For instance, considering Table \ref{output}, even if the ``Dashboard Installation'' \texttt{Process\-Element} is not affected directly by a cyber security risk, through our approach we can uncover that there is a risk over that element, derived through its \texttt{Followed\-By} dependency with ``DoorDisassembly'' object (see Fig. \ref{subgraph}).

\section{Related Work}
\label{rel}

There is a lively and quite growing interest in risk propagation (or inheritance) techniques and their applications for risk assessment. However, the research on the exploitation of ontologies for process-aware risk propagation is much more restricted.

In this focused area of research, for what concerns the usage of ontologies, the seminal work in \cite{ou2006scalable} presents a solution that is relevant to our proposal. The main contribution here is to show how an \textit{ad hoc} graph, i.e., the attack graph, generated through the \textit{MulVAL} approach, can be used to encode logical reasoning methods and inference rules gathered by cyber-security experts. Such a graph can then be exploited to quantify the cyber-security risk of a system and infer what steps an adversary can perform to reach a pre-defined target. In the same line of research, the work in \cite{cao2019ontology} develops a risk propagation ontology-based \textit{bayesian network} (BN) model to measure dynamic \textit{supply chain risk} (SCR) propagation, with the main purpose of quantitatively assessing the impact of dynamic risk propagation within and between integrated firms in global fresh produce supply chains. Similarly, the work presented in \cite{de2013using} provides an ontology, created by leveraging the knowledge of security experts, to apply risk propagation over a given attack graph, by also supporting a score calculation. 

These and similar approaches do not yet consider the impact of risk at different levels of abstraction. For instance, the original MulVAL approach \cite{ou2006scalable} focuses on inferring the potential movements of the attacker within an infrastructure layer, but cannot be used to quantify the risk or propagate it across physical components and business processes. In this regard, the first work that emphasizes the need to analyze this aspect and to apply risk propagation from physical structures to internal economic processes within an organization was the one presented in \cite{musman2011computing}.

A few years after, moved by the same problem, the authors of \cite{cao2018assessing} presented an approach to assessing the cyber-attack impact on business processes, by generating an interconnected graph of the dependencies between vulnerabilities on hosts, relations between services to hosts, and tasks to services. The authors encoded the dependencies with a Datalog\footnote{\url{https://docs.racket-lang.org/datalog/}} model, collecting a large set of facts and rules, extracted via MulVAL. In this work, the method proposed to calculate the impact score by propagating the impact generated by the vulnerabilities to the impacted hosts, services, and tasks, is similar to ours. However, in our approach, the knowledge component is not designed to cover the cyber-security domain only, thus making our solution different in terms of generality.

Similarly, the authors in \cite{haque2020modeling} used an approach to assess the mission impact of cyber-attacks on energy delivery systems, cyber-physical systems, and enterprise applications, respectively. Finally, recent research by \cite{gonzalez2021quantifying} presented a method to propagate the risk within a network of business processes and IT services, by transforming the operational risk over the underlying IT services into a financial risk over their related business processes. Still, these two last reported works, even if going towards a process-aware approach to risk assessment, lack in considering the phenomenon from a cross-domain perspective, i.e., independently from the context for which they were developed (i.e., cyber-security and IT services).

\section{Conclusion and Perspectives}
\label{cons}

This paper presents an application that leverages the combination of \textit{i)} an ontology, encoding a set of rules to be used for representing the risk dependencies within a system composed of objects and processes and \textit{ii)} a method to calculate the propagation of risk within the represented system. By doing so we move towards the development of a process-aware risk-propagation approach that is aimed at facilitating the assessment of risk propagation between processes and objects with different levels of abstraction.  

Among the potential paths of research opened up by our results, we envision a series of future perspectives. A first perspective concerns the extension of the Risk-Process ontology. We plan to leverage previous work on risk and value modeling \cite{sales2018common} and provide a well-founded ontology for process-aware risk propagation. A second perspective concerns the implementation of different algorithms for the calculation and the propagation of risk, considering also different kinds of dependencies between the elements at risk in the graph. A third perspective is to apply the approach over multiple domains (e.g., \textit{finance} and \textit{healthcare}), where risk plays a central role and pave the way for multiple case studies.

\begin{acks}
This work was done in collaboration with Accenture Labs, Israel.
\end{acks}

\bibliographystyle{ACM-Reference-Format}
\bibliography{sample-bibliography} 


\begin{thebibliography}{22}


\ifx \showCODEN    \undefined \def \showCODEN     #1{\unskip}     \fi
\ifx \showDOI      \undefined \def \showDOI       #1{#1}\fi
\ifx \showISBNx    \undefined \def \showISBNx     #1{\unskip}     \fi
\ifx \showISBNxiii \undefined \def \showISBNxiii  #1{\unskip}     \fi
\ifx \showISSN     \undefined \def \showISSN      #1{\unskip}     \fi
\ifx \showLCCN     \undefined \def \showLCCN      #1{\unskip}     \fi
\ifx \shownote     \undefined \def \shownote      #1{#1}          \fi
\ifx \showarticletitle \undefined \def \showarticletitle #1{#1}   \fi
\ifx \showURL      \undefined \def \showURL       {\relax}        \fi
\providecommand\bibfield[2]{#2}
\providecommand\bibinfo[2]{#2}
\providecommand\natexlab[1]{#1}
\providecommand\showeprint[2][]{arXiv:#2}

\bibitem[\protect\citeauthoryear{Alexander, Belisle, and Steele}{Alexander
  et~al\mbox{.}}{2020}]%
        {alexander2020mitre}
\bibfield{author}{\bibinfo{person}{Otis Alexander}, \bibinfo{person}{Misha
  Belisle}, {and} \bibinfo{person}{Jacob Steele}.}
  \bibinfo{year}{2020}\natexlab{}.
\newblock \showarticletitle{MITRE ATT\&CK for Industrial Control Systems:
  Design and Philosophy}.
\newblock \bibinfo{journal}{\emph{The MITRE Corporation: Bedford, MA, USA}}
  (\bibinfo{year}{2020}).
\newblock


\bibitem[\protect\citeauthoryear{Awerbuch}{Awerbuch}{1985}]%
        {awerbuch1985new}
\bibfield{author}{\bibinfo{person}{Baruch Awerbuch}.}
  \bibinfo{year}{1985}\natexlab{}.
\newblock \showarticletitle{A new distributed depth-first-search algorithm}.
\newblock \bibinfo{journal}{\emph{Inform. Process. Lett.}}
  \bibinfo{volume}{20}, \bibinfo{number}{3} (\bibinfo{year}{1985}),
  \bibinfo{pages}{147--150}.
\newblock


\bibitem[\protect\citeauthoryear{Cao, Yuan, Singhal, Liu, Sun, and Zhu}{Cao
  et~al\mbox{.}}{2018a}]%
        {kerschbaum_assessing_2018}
\bibfield{author}{\bibinfo{person}{Chen Cao}, \bibinfo{person}{Lun-Pin Yuan},
  \bibinfo{person}{Anoop Singhal}, \bibinfo{person}{Peng Liu},
  \bibinfo{person}{Xiaoyan Sun}, {and} \bibinfo{person}{Sencun Zhu}.}
  \bibinfo{year}{2018}\natexlab{a}.
\newblock \showarticletitle{Assessing {Attack} {Impact} on {Business}
  {Processes} by {Interconnecting} {Attack} {Graphs} and {Entity} {Dependency}
  {Graphs}}.
\newblock In \bibinfo{booktitle}{\emph{Data and {Applications} {Security} and
  {Privacy} {XXXII}}}, \bibfield{editor}{\bibinfo{person}{Florian Kerschbaum}
  {and} \bibinfo{person}{Stefano Paraboschi}} (Eds.).
  Vol.~\bibinfo{volume}{10980}. \bibinfo{publisher}{Springer International
  Publishing}, \bibinfo{address}{Cham}, \bibinfo{pages}{330--348}.
\newblock
\showISBNx{978-3-319-95728-9 978-3-319-95729-6}
\urldef\tempurl%
\url{https://doi.org/10.1007/978-3-319-95729-6_21}
\showDOI{\tempurl}


\bibitem[\protect\citeauthoryear{Cao, Yuan, Singhal, Liu, Sun, and Zhu}{Cao
  et~al\mbox{.}}{2018b}]%
        {cao2018assessing}
\bibfield{author}{\bibinfo{person}{Chen Cao}, \bibinfo{person}{Lun-Pin Yuan},
  \bibinfo{person}{Anoop Singhal}, \bibinfo{person}{Peng Liu},
  \bibinfo{person}{Xiaoyan Sun}, {and} \bibinfo{person}{Sencun Zhu}.}
  \bibinfo{year}{2018}\natexlab{b}.
\newblock \showarticletitle{Assessing attack impact on business processes by
  interconnecting attack graphs and entity dependency graphs}. In
  \bibinfo{booktitle}{\emph{IFIP Annual Conference on Data and Applications
  Security and Privacy}}. Springer, \bibinfo{pages}{330--348}.
\newblock


\bibitem[\protect\citeauthoryear{Cao, Bryceson, and Hine}{Cao
  et~al\mbox{.}}{2019}]%
        {cao2019ontology}
\bibfield{author}{\bibinfo{person}{Shoufeng Cao}, \bibinfo{person}{Kim
  Bryceson}, {and} \bibinfo{person}{Damian Hine}.}
  \bibinfo{year}{2019}\natexlab{}.
\newblock \showarticletitle{An Ontology-based Bayesian network modelling for
  supply chain risk propagation}.
\newblock \bibinfo{journal}{\emph{Industrial Management \& Data Systems}}
  (\bibinfo{year}{2019}).
\newblock


\bibitem[\protect\citeauthoryear{Choudhary, Singh, Schoenherr, and
  Ramkumar}{Choudhary et~al\mbox{.}}{2022}]%
        {choudhary2022risk}
\bibfield{author}{\bibinfo{person}{Nishat~Alam Choudhary},
  \bibinfo{person}{Shalabh Singh}, \bibinfo{person}{Tobias Schoenherr}, {and}
  \bibinfo{person}{M Ramkumar}.} \bibinfo{year}{2022}\natexlab{}.
\newblock \showarticletitle{Risk assessment in supply chains: a
  state-of-the-art review of methodologies and their applications}.
\newblock \bibinfo{journal}{\emph{Annals of Operations Research}}
  (\bibinfo{year}{2022}), \bibinfo{pages}{1--43}.
\newblock


\bibitem[\protect\citeauthoryear{Das, Srinivasan, Perry, Chong, and
  Banerjee}{Das et~al\mbox{.}}{2014}]%
        {das2014tale}
\bibfield{author}{\bibinfo{person}{Souripriya Das},
  \bibinfo{person}{Jagannathan Srinivasan}, \bibinfo{person}{Matthew Perry},
  \bibinfo{person}{Eugene~Inseok Chong}, {and} \bibinfo{person}{Jayanta
  Banerjee}.} \bibinfo{year}{2014}\natexlab{}.
\newblock \showarticletitle{A Tale of Two Graphs: Property Graphs as RDF in
  Oracle.}. In \bibinfo{booktitle}{\emph{EDBT}}. \bibinfo{pages}{762--773}.
\newblock


\bibitem[\protect\citeauthoryear{de~Barros~Barreto, da~Costa, and
  Yano}{de~Barros~Barreto et~al\mbox{.}}{2013}]%
        {de2013using}
\bibfield{author}{\bibinfo{person}{Alexandre de Barros~Barreto},
  \bibinfo{person}{Paulo Cesar~G da Costa}, {and}
  \bibinfo{person}{Edgar~Toshiro Yano}.} \bibinfo{year}{2013}\natexlab{}.
\newblock \showarticletitle{Using a Semantic Approach to Cyber Impact
  Assessment.}. In \bibinfo{booktitle}{\emph{STIDS}}.
  \bibinfo{pages}{101--108}.
\newblock


\bibitem[\protect\citeauthoryear{Dumas, La~Rosa, Mendling, Reijers,
  et~al\mbox{.}}{Dumas et~al\mbox{.}}{2013}]%
        {dumas2013fundamentals}
\bibfield{author}{\bibinfo{person}{Marlon Dumas}, \bibinfo{person}{Marcello
  La~Rosa}, \bibinfo{person}{Jan Mendling}, \bibinfo{person}{Hajo~A Reijers},
  {et~al\mbox{.}}} \bibinfo{year}{2013}\natexlab{}.
\newblock \bibinfo{booktitle}{\emph{Fundamentals of business process
  management}}. Vol.~\bibinfo{volume}{1}.
\newblock \bibinfo{publisher}{Springer}.
\newblock


\bibitem[\protect\citeauthoryear{Engelberg}{Engelberg}{2022}]%
        {engelberg2022process}
\bibfield{author}{\bibinfo{person}{Gal Engelberg}.}
  \bibinfo{year}{2022}\natexlab{}.
\newblock \showarticletitle{Process-Aware Attack-Graphs for Risk Quantification
  and Mitigation in Industrial Infrastructures}.
\newblock  (\bibinfo{year}{2022}).
\newblock


\bibitem[\protect\citeauthoryear{Fenrich}{Fenrich}{2008}]%
        {fenrich2008securing}
\bibfield{author}{\bibinfo{person}{Kim Fenrich}.}
  \bibinfo{year}{2008}\natexlab{}.
\newblock \showarticletitle{Securing your control system: the" CIA triad" is a
  widely used benchmark for evaluating information system security
  effectiveness}.
\newblock \bibinfo{journal}{\emph{Power Engineering}} \bibinfo{volume}{112},
  \bibinfo{number}{2} (\bibinfo{year}{2008}), \bibinfo{pages}{44--49}.
\newblock


\bibitem[\protect\citeauthoryear{Gonz{\'a}lez-Rojas, Castro, and
  Lesmes}{Gonz{\'a}lez-Rojas et~al\mbox{.}}{2021}]%
        {gonzalez2021quantifying}
\bibfield{author}{\bibinfo{person}{Oscar Gonz{\'a}lez-Rojas},
  \bibinfo{person}{Nicol{\'a}s Castro}, {and} \bibinfo{person}{Sebastian
  Lesmes}.} \bibinfo{year}{2021}\natexlab{}.
\newblock \showarticletitle{Quantifying Risk Propagation Within a Network of
  Business Processes and IT Services}.
\newblock \bibinfo{journal}{\emph{Business \& Information Systems Engineering}}
  \bibinfo{volume}{63}, \bibinfo{number}{2} (\bibinfo{year}{2021}),
  \bibinfo{pages}{129--143}.
\newblock


\bibitem[\protect\citeauthoryear{Hadar, Kravchenko, and Basovskiy}{Hadar
  et~al\mbox{.}}{2020}]%
        {hadar_cyber_2020}
\bibfield{author}{\bibinfo{person}{Ethan Hadar}, \bibinfo{person}{Dmitry
  Kravchenko}, {and} \bibinfo{person}{Alexander Basovskiy}.}
  \bibinfo{year}{2020}\natexlab{}.
\newblock \showarticletitle{Cyber {Digital} {Twin} {Simulator} for {Automatic}
  {Gathering} and {Prioritization} of {Security} {Controls}’ {Requirements}}.
  In \bibinfo{booktitle}{\emph{2020 {IEEE} 28th {International} {Requirements}
  {Engineering} {Conference} ({RE})}}. \bibinfo{pages}{250--259}.
\newblock
\urldef\tempurl%
\url{https://doi.org/10.1109/RE48521.2020.00035}
\showDOI{\tempurl}
\newblock
\shownote{ISSN: 2332-6441.}


\bibitem[\protect\citeauthoryear{Haque, Shetty, Kamhoua, and Gold}{Haque
  et~al\mbox{.}}{2020}]%
        {haque2020modeling}
\bibfield{author}{\bibinfo{person}{Md~Ariful Haque}, \bibinfo{person}{Sachin
  Shetty}, \bibinfo{person}{Charles~A Kamhoua}, {and} \bibinfo{person}{Kimberly
  Gold}.} \bibinfo{year}{2020}\natexlab{}.
\newblock \showarticletitle{Modeling Mission Impact of Cyber Attacks on Energy
  Delivery Systems}. In \bibinfo{booktitle}{\emph{International Conference on
  Security and Privacy in Communication Systems}}. Springer,
  \bibinfo{pages}{41--61}.
\newblock


\bibitem[\protect\citeauthoryear{Hevner, March, Park, and Ram}{Hevner
  et~al\mbox{.}}{2004}]%
        {hevner2004design}
\bibfield{author}{\bibinfo{person}{Alan~R Hevner}, \bibinfo{person}{Salvatore~T
  March}, \bibinfo{person}{Jinsoo Park}, {and} \bibinfo{person}{Sudha Ram}.}
  \bibinfo{year}{2004}\natexlab{}.
\newblock \showarticletitle{Design science in information systems research}.
\newblock \bibinfo{journal}{\emph{MIS quarterly}} (\bibinfo{year}{2004}),
  \bibinfo{pages}{75--105}.
\newblock


\bibitem[\protect\citeauthoryear{ISO 31000}{ISO 31000}{2018}]%
        {ISO2018}
ISO 31000 \bibinfo{year}{2018}\natexlab{}.
\newblock \bibinfo{booktitle}{\emph{{Risk Management - Guidelines}}}.
\newblock \bibinfo{type}{Standard}.
\newblock


\bibitem[\protect\citeauthoryear{Jiang, Wen, Yu, Xiang, and Zhou}{Jiang
  et~al\mbox{.}}{2016}]%
        {jiang2016identifying}
\bibfield{author}{\bibinfo{person}{Jiaojiao Jiang}, \bibinfo{person}{Sheng
  Wen}, \bibinfo{person}{Shui Yu}, \bibinfo{person}{Yang Xiang}, {and}
  \bibinfo{person}{Wanlei Zhou}.} \bibinfo{year}{2016}\natexlab{}.
\newblock \showarticletitle{Identifying propagation sources in networks:
  State-of-the-art and comparative studies}.
\newblock \bibinfo{journal}{\emph{IEEE Communications Surveys \& Tutorials}}
  \bibinfo{volume}{19}, \bibinfo{number}{1} (\bibinfo{year}{2016}),
  \bibinfo{pages}{465--481}.
\newblock


\bibitem[\protect\citeauthoryear{Kavallieratos, Spathoulas, and
  Katsikas}{Kavallieratos et~al\mbox{.}}{2021}]%
        {kavallieratos2021cyber}
\bibfield{author}{\bibinfo{person}{Georgios Kavallieratos},
  \bibinfo{person}{Georgios Spathoulas}, {and} \bibinfo{person}{Sokratis
  Katsikas}.} \bibinfo{year}{2021}\natexlab{}.
\newblock \showarticletitle{Cyber risk propagation and optimal selection of
  cybersecurity controls for complex cyberphysical systems}.
\newblock \bibinfo{journal}{\emph{Sensors}} \bibinfo{volume}{21},
  \bibinfo{number}{5} (\bibinfo{year}{2021}), \bibinfo{pages}{1691}.
\newblock


\bibitem[\protect\citeauthoryear{Musman, Tanner, Temin, Elsaesser, and
  Loren}{Musman et~al\mbox{.}}{2011}]%
        {musman2011computing}
\bibfield{author}{\bibinfo{person}{Scott Musman}, \bibinfo{person}{Mike
  Tanner}, \bibinfo{person}{Aaron Temin}, \bibinfo{person}{Evan Elsaesser},
  {and} \bibinfo{person}{Lewis Loren}.} \bibinfo{year}{2011}\natexlab{}.
\newblock \showarticletitle{Computing the impact of cyber attacks on complex
  missions}. In \bibinfo{booktitle}{\emph{2011 IEEE International Systems
  Conference}}. IEEE, \bibinfo{pages}{46--51}.
\newblock


\bibitem[\protect\citeauthoryear{Oliveira, Sales, Baratella, Fumagalli, and
  Guizzardi}{Oliveira et~al\mbox{.}}{2022}]%
        {oliveiraontology}
\bibfield{author}{\bibinfo{person}{{\'I}talo Oliveira},
  \bibinfo{person}{Tiago~Prince Sales}, \bibinfo{person}{Riccardo Baratella},
  \bibinfo{person}{Mattia Fumagalli}, {and} \bibinfo{person}{Giancarlo
  Guizzardi}.} \bibinfo{year}{2022}\natexlab{}.
\newblock \showarticletitle{An Ontology of Security from a Risk Treatment
  Perspective}. In \bibinfo{booktitle}{\emph{International conference on
  conceptual modeling}}. Springer.
\newblock


\bibitem[\protect\citeauthoryear{Ou, Boyer, and McQueen}{Ou
  et~al\mbox{.}}{2006}]%
        {ou2006scalable}
\bibfield{author}{\bibinfo{person}{Xinming Ou}, \bibinfo{person}{Wayne~F
  Boyer}, {and} \bibinfo{person}{Miles~A McQueen}.}
  \bibinfo{year}{2006}\natexlab{}.
\newblock \showarticletitle{A scalable approach to attack graph generation}. In
  \bibinfo{booktitle}{\emph{Proceedings of the 13th ACM conference on Computer
  and communications security}}. \bibinfo{pages}{336--345}.
\newblock


\bibitem[\protect\citeauthoryear{Sales, Bai{\~a}o, Guizzardi, Almeida, Guarino,
  and Mylopoulos}{Sales et~al\mbox{.}}{2018}]%
        {sales2018common}
\bibfield{author}{\bibinfo{person}{Tiago~Prince Sales},
  \bibinfo{person}{Fernanda Bai{\~a}o}, \bibinfo{person}{Giancarlo Guizzardi},
  \bibinfo{person}{Jo{\~a}o Paulo~A Almeida}, \bibinfo{person}{Nicola Guarino},
  {and} \bibinfo{person}{John Mylopoulos}.} \bibinfo{year}{2018}\natexlab{}.
\newblock \showarticletitle{The common ontology of value and risk}. In
  \bibinfo{booktitle}{\emph{International conference on conceptual modeling}}.
  Springer, \bibinfo{pages}{121--135}.
\newblock


\end{thebibliography}

\end{document}